\documentstyle[epsf,11pt]{article}
\begin{document}

\begin{titlepage}

\begin{flushright}
   TPR-97-12
\end{flushright}

\vspace{1.5cm}

\begin{center}
{\Large\bf Fine-Tuning Two-Particle Interferometry\protect\\ 
 I: Effects from Temperature Gradients in the Source}
\end{center}
  
\vspace{1cm}

\begin{center}
 {\large Boris Tom\'a\v{s}ik and Ulrich Heinz}
\end{center}

\vspace{0.5cm}

\begin{center}
   {\it    
   Institut f\"ur Theoretische Physik, Universit\"at Regensburg,\\
   D-93040 Regensburg, Germany}

\vspace{2.5cm}

{September 24, 1997}
\end{center}

\vspace{1.1cm}

\abstract
{A comprehensive model study of Bose-Einstein correlation
radii in heavy ion collisions is presented. The starting point
is a longitudinally and transversally expanding fireball,
represented at freeze-out by an azimuthally symmetric emission 
function. The freeze-out temperature is allowed to feature transverse 
and temporal gradients. Their effects on the correlation radii are 
studied. In particular, we evaluate numerically their dependence on
the transverse mass of the particle pairs and check a 
recent suggestion, based on analytical approximations, that for
certain reasonable source parameters all three correlation radii
satisfy simultaneously a $1/\sqrt{M_\perp}$ scaling.
}
\endabstract
\vspace{2.5cm}

\end{titlepage}

\section{Introduction}
\label{intro}

Recently, two-particle intensity interferometry, exploiting the effects 
of Bose-Einstein (BE) symmetrization on the two-particle momentum spectra,
has been developed into a powerful tool for measuring not only 
the space-time dimensions of the particle emitting object, 
but also its dynamical state at particle freeze-out. Modern 
Bose-Einstein interferometry thus goes far beyond the original
work by Hanbury Brown and Twiss, who introduced and successfully
demonstrated photon intensity interferometry for static sources 
in astrophysics \cite{HBT54}, and the pioneering paper by Goldhaber,
Goldhaber, Lee, and Pais \cite{GGLP} who first exploited similar ideas
in particle physics. Good reviews of the basic theoretical and experimental 
techniques can be found in~\cite{GKW79,L89,BGJ90,Z93,P95} while the
more recent theoretical developments are summarized in a series of lectures
published in~\cite{H96}. 

Collective dynamics of the source leads to a characteristic dependence 
of the two-particle correlation function on the total momentum of the 
particle pair. The width of the correlator as a function of the relative 
momentum of the two particles measures the size of the ``regions of 
homogeneity'' \cite{MS88} inside the source across which the momentum
distribution changes sufficiently little to guarantee measurable 
rates for pairs with similar momenta of the two particles. For rapidly
expanding sources widely separated points do not emit particles of
similar momenta; the homogeneity regions seen by BE correlation
function thus form only a fraction of the whole source. Moreover, the
momentum of a particle defines the homogeneity region from which
it is coming; we will call the homogeneity region which contributes to
the emission of particles with a given momentum the ``effective
source'' for such particles. 

The size parameters of the ``effective source'' (which are measured by the
correlator and which we will thus call ``correlation radii'' or simply 
``HBT radii'') are related to the magnitude of the velocity gradients 
in the source, multiplied with a thermal smearing factor $\sqrt{T/M_\perp}$ 
originating from the local momentum distribution \cite{CSH95a,CSH95b,WSH96}. 
They are thus affected by both the expansion velocity profile of the 
source and by variations of the width (``temperature'') of the local 
momentum distributions across the source \cite{CSH95b,CL95,CL96a,CL96b,AS96}. 
Another physical mechanism which can, even in the absence of flow and
temperature gradients, lead to dramatic differences between the 
``effective source'' of particles with a fixed momentum $K$ and 
the momentum-integrated distribution of source points in space-time 
is strong reabsorption or rescattering of the particles in the interior 
of the source; this leads to a surface-dominated distribution of last 
interaction points with a strong preference for outward-directed 
momenta. Such ``opaque sources'' and their effects on the correlation radii
have recently been discussed in~\cite{HV96,HV97} on a semianalytic 
level using simple models. Our investigations on this subject will be 
published in the next paper~\cite{th2}.

While the basic relations between such source features as mentioned above 
and the properties of the measurable two-particle correlation functions 
appear to be qualitatively understood, this is not sufficient for a 
quantitative interpretation of two-particle correlation data and a complete 
reconstruction of the source from the measurements. Quantitative 
numerical investigations, including comprehensive parameter studies, 
so far exist mostly for transparent expanding sources with constant 
freeze-out temperature \cite{WSH96,HTWW96a,WHTW96,THWW97}, while for 
sources with temperature gradients some numerical checks of the approximate 
analytical approximations developed in \cite{CL95,CL96a,CL96b} have
been performed in~\cite{CLL96,CL96c}. In \cite{CLL96} the validity
range of the analytical formulae given in~\cite{CL96a} was
determined. A comparison with experimental data from 200 $A$ GeV/$c$ S+Pb 
collisions \cite{NA44} in~\cite{CL96c} revealed, however, that
the required source parameters lie outside this range of validity.
A quantitative comparison with data thus requires numerical studies.

In this paper  we present
a numerical parameter study of the influence of flow and temperature
gradients on the correlation radii. One goal of this investigation
is to settle a question which was left open in~\cite{CLL96,CL96c}:
based on a series of approximations, the authors of~\cite{CL96a}
had found a common $1/\sqrt{M_\perp}$ scaling law for all three 
HBT radii $R_s$, $R_o$, $R_l$ in a Cartesian parametrisation of the 
correlator. This was desirable in view of the S+Pb data of~\cite{NA44} 
where agreement with such a common scaling law was 
reported (although supported only by three data points in each
Cartesian direction). The data 
of~\cite{NA35,AlberPhD,Alber95,NA49QM96,Stefan,Harry} from S+S,
S+Ag, S+Au, and Pb+Pb collisions, on the other hand, all  
show a much stronger $M_\perp$-dependence of the longitudinal HBT radii
compared to the transverse ones. This is in qualitative
agreement with the theoretical studies in~\cite{WSH96,HTWW96a,WHTW96}
which, for sources with constant temperature and boost-invariant 
longitudinal, but weaker transverse expansion, show a weaker 
$M_\perp$-dependence for $R_s$ than for $R_l$. In \cite{WSH96,WHTW96} 
it was argued that, if one fits the $M_\perp$-dependence of the 
HBT radii with a negative power law, $R_i(M_\perp) \sim 
M_\perp^{-\alpha_i}$ \cite{AlberPhD,Alber95,Harry}, the power $\alpha_i$ 
itself should be proportional to the rate of expansion in direction $i$; 
this feature cannot be obtained within the simple saddle point 
approximation used in~\cite{CSH95a,CSH95b,CL95,CL96a,AS96,CNH95}. 
We show here that, within the phenomenologically allowed parameter range,
this remains true even in the presence of temperature gradients.

In contrast to the work of~\cite{CLL96,CL96c} we also compute the
HBT radii in the Yano-Koonin-Podgoretski\u\i \ (YKP) parametrisation
of the correlation function \cite{YK78,P83,CNH95} which have a more 
straightforward and simpler interpretation in space-time 
\cite{CNH95,WHTW96} than the HBT parameters from the Cartesian
(Pratt-Bertsch-Chapman) parametrisation \cite{CSH95a,CSH95b}. As far
as we know, the effects of temperature gradients on the YKP parameters
have not been studied before. 

We want to stress that resonance decays are not addressed in this paper.
It is known that they can strongly affect the correlation function 
\cite{Bolz93,Schlei96,Ornik96,Heis96,CLZ96,reson1} in which case a
Gaussian parametrisation as employed here becomes questionable 
\cite{reson1,reson2,reson3}. Thus, while our calculations refer 
both to pion and kaon correlations, only the kaon results and those 
for high-$K_\perp$ pions (where resonance decays can be neglected) 
can be directly related to data. For low-$K_\perp$ pion pairs our
results show the features of the contribution from directly emitted
pions to the correlator.

\section{Formalism}
\label{formalism}

For chaotic sources the two-particle correlation function is in 
very good approximation given by the formula \cite{S73,P84,CH94}
\begin{equation}
C(q,K) \simeq 1 +
   \frac {{\left | \int d^4x \, S(x,K) \, e^{iq.x} \right |}^2}
   {{\left | \int d^4x \, S(x,K) \, \right | }^2 } \, .
\label{2.1}
\end{equation}
Here $S(x,K)$ is the emission function (single-particle Wigner phase-space
density) of the source \cite{S73,P84,CH94}, and $K=\frac{1}{2}(p_1+p_2)$
(with $p_{1,2}$ on-shell) is the average pair momentum while
$q = p_1 - p_2$ denotes the momentum difference between the two particles.

One usually parameterizes the correlation function by a Gaussian in $q$
\cite{HTWW96a,WHTW96}. Since $q$ satisfies the on-shell constraint 
$q\cdot K=0$, only three of its four components are independent. This 
leaves room for various (mathematically equivalent) Gaussian 
parametrisations, using different sets of independent $q$-components.
Here we will consider only azimuthally symmetric sources and evaluate 
the parameters (HBT radii) of the Cartesian and 
of the Yano-Koonin-Podgoretski\u\i \ (YKP) parametrisations, in the 
commonly used coordinate system where the $z$ axis defines the beam 
direction and ${\bf K}$ lies in the $x-z$ plane. The $x$ axis is 
customarily labeled as $o$ (for {\it outward}), $y$ as $s$ (for 
{\it sideward}), and $z$ as $l$ (for {\it longitudinal}).

The Cartesian parametrisation of the correlator employs the three 
spatial components of $q$ in the form \cite{CSH95a,CSH95b} 
\begin{equation}
    C({\bf q},{\bf K})
    = 1 + \exp\left[ - R_s^2({\bf K}) q_s^2 - R_o^2({\bf K}) q_o^2
                     - R_l^2({\bf K}) q_l^2 - 2 R_{ol}^2({\bf K}) q_o q_l
              \right] \, .
\label{2.8}
\end{equation}
The four ${\bf K}$-dependent parameters (HBT radii) are then given
by linear combinations of the space-time variances of the effective
source of particles with momentum\footnote{We denote the dependence on
  the pair momentum alternatively by ${\bf K}$ or $K$ where in the
  latter case the on-shell approximation $K^0 = \sqrt{ m^2 + {\bf
      K}^2}$ is implied.} $K$ \cite{CSH95a,CSH95b,HB95}:
\begin{eqnarray}
   R_s^2({\bf K}) &=& \langle \tilde{y}^2 \rangle \, ,
 \label{2.9a}\\
   R_o^2({\bf K}) &=&
   \langle (\tilde{x} - \beta_\perp \tilde t)^2 \rangle \, ,
 \label{2.9b}\\
   R_l^2({\bf K}) &=&
   \langle (\tilde{z} - \beta_l \tilde t)^2 \rangle \, ,
 \label{2.9c}\\
   R_{ol}^2({\bf K}) &=&
   \langle (\tilde{x} - \beta_\perp \tilde t)
           (\tilde{z} - \beta_l \tilde t) \rangle \, .
 \label{2.9d}
\end{eqnarray}
The angular brackets denote space-time averages taken with the emission 
function $S(x,K)$ of the effective source at momentum $K$, and coordinates
with a tilde are measured relative to the center of the effective source:
$\tilde x_\mu = x_\mu - \bar x_\mu \, , \bar x_\mu = \langle x_\mu \rangle$
\cite{HTWW96a}.

In the LCMS (Longitudinally CoMoving System \cite{CP91}), where
$\beta_l = K_l = 0$, $R_l^2$ and $R_{ol}^2$ simplify to
\begin{eqnarray}
R_l^2 & = & \langle {\tilde z}^2 \rangle \, ,
\label{2.10a} \\
R_{ol}^2 & = & \langle (\tilde x - \beta_\perp \tilde t)\, \tilde z
\rangle \, .
\label{2.10b}
\end{eqnarray}
For further discussions of these expressions see \cite{CSH95a,CSH95b}.

The YKP parametrisation employs the components $q^0$, $q_\perp = 
\sqrt{q_o^2 + q_s^2}$, and $q_l$ in the form \cite{CNH95,HTWW96a,WHTW96}
\begin{equation}
   C({\bf q},{\bf K}) =
       1 +  \exp\left[ - R_\perp^2({\bf K})\, q_{\perp}^2
                       - R_\parallel^2({\bf K}) \left( q_l^2{-}(q^0)^2 \right)
                       - \left( R_0^2({\bf K}){+}R_\parallel^2({\bf K})\right)
                         \left(q{\cdot}U({\bf K})\right)^2
                \right] .
\label{2.12}
\end{equation}
The YKP radii $R_\perp$, $R_\parallel$, and $R_0$ are invariant under 
longitudinal boosts of the measurement frame; they measure (in some 
approximation) the transverse, longitudinal, and temporal extension 
of the effective source in its longitudinal rest frame (called Yano-Koonin
(YK) frame) \cite{CNH95,HTWW96a,WHTW96}. This frame moves with the 
YK velocity $v({\bf K})$, defined by the fourth fit parameter 
$U({\bf K})$ via
\begin{equation}
U({\bf K}) = \gamma ({\bf K})\,(1,0,0,v({\bf K})), \qquad 
\gamma ({\bf K}) = \frac 1{\sqrt{1 - v^2 ({\bf K})}} \, .
\label{2.13}
\end{equation}
The YKP parameters can be calculated from the emission function through
relations similar to (\ref{2.9a})-(\ref{2.9d}): Defining
\cite{HTWW96a,WHTW96} 
\begin{eqnarray}
 \label{2.14a}
   A &=& \left\langle \left( \tilde t
         - {\tilde \xi\over \beta_\perp} \right)^2 \right\rangle \, ,
 \\
 \label{2.14b}
   B &=&  \left\langle \left( \tilde z
         - {\beta_l\over \beta_\perp} \tilde \xi \right)^2 \right\rangle
   \, ,
 \\
 \label{2.14c}
   C &=& \left\langle \left( \tilde t - {\tilde \xi\over \beta_\perp} \right)
                      \left( \tilde z - {\beta_l\over \beta_\perp}
                             \tilde \xi \right) \right\rangle \, ,
\end{eqnarray}
with $\tilde \xi \equiv \tilde x + i \tilde y$ such that $\langle \tilde 
\xi^2 \rangle = \langle \tilde x^2 - \tilde y^2 \rangle$, they are given 
by\footnote{Please note that the first expressions given on the r.h.s of 
Eqs.~(19b,c) in~\protect\cite{HTWW96a} are only valid if $A+B \geq 0$.}
\begin{eqnarray}
 \label{2.15a}
   v &=& {A+B\over 2C} \left( 1 - \sqrt{1 - \left({2C\over A+B}\right)^2}
                       \right) \, ,
 \\
 \label{2.15b}
   R_\parallel^2 &=& B - v\, C \, ,
\\
 \label{2.15c}
   R_0^2 &=& A - v\, C\, ,
\\
 \label{2.15d}
   R_{\perp}^2 &=& {\langle{ \tilde{y}^2 }\rangle}\, .
\end{eqnarray}
Note that the Yano-Koonin velocity $v$, and thus $R_\parallel$ and $R_0$,
are well-defined only for 
effective sources with $(A+B)^2 - 4C^2 \ge 0$. It will be shown in 
\cite{th2} that  this condition 
can be violated in particular for opaque sources. In the limit $C\to 0$
the YK velocity $v$ vanishes such that this condition defines the YK 
frame. In this frame $R_\parallel^2 = B$ and $R_0^2 = A$.

As a side remark, we would like to note that the quantities $A,B,C$ from 
Eqs.~(\ref{2.14a})-(\ref{2.14c}) are identical with the radius parameters
introduced in~\cite{THWW97,reson1,reson3}, where a different 
Cartesian parametrisation based on the same $q$-components as in the 
YKP case was used:
\begin{equation}
\label{2.20}
C({\bf q},{\bf K}) = 1 + \exp  [- R_\perp^2({\bf K})q_\perp^2 -
   R_z^2({\bf K}) q_l^2 - R_t^2({\bf K})(q^0)^2 -
   2\, R_{zt}^2({\bf K}) q_l q^0 ] \, .
\end{equation}
One easily shows
\begin{equation}
\label{2.21}
R^2_t = A \, ,\quad R^2_z = B \, , \quad R^2_{zt} = - C \, .
\end{equation}
Clearly, these parameters are not boost-invariant and thus
don't lead directly to a simple space-time interpretation of
the correlator. Their usefulness is rather of technical nature
in the actual data fitting procedure \cite{THWW97,reson1,reson3}
and, of course, the YKP parameters can be reconstructed from them
via Eqs.~(\ref{2.15a})-(\ref{2.15c}).

Our calculations in the following sections of the correlation radii 
will be based on the expressions (\ref{2.9a})-(\ref{2.10b}) and 
(\ref{2.14a})-(\ref{2.15d}).

\section{Hydrodynamic parametrisation of the source}
\label{hydromodel}

For our studies we use slightly modified (see Appendix~\ref{tam_model})
model of~\cite{CL96a}:
\begin{eqnarray}
   S(x,K) \, d^4x & = & {M_\perp \cosh(\eta-Y) \over
            (2\pi)^3 }
        \; \exp \left[- {K \cdot u(x) \over T(x)}\right]
        \; \exp \left[ - {r^2 \over 2 R^2}
            - {{(\eta- \eta_0)}^2 \over 2 (\Delta \eta)^2}
           \right]         \nonumber \\ &  & \times 
            \frac{\tau \, d\tau}{\sqrt{2\pi(\Delta \tau)^2}}  \;
            \exp \left[- {(\tau-\tau_0)^2 \over 2(\Delta \tau)^2}
           \right ] \, d\eta \, r\, dr\, d\phi.
\label{3.1}
\end{eqnarray}
This ``emission function'' parametrizes the distribution of points of last 
interaction in the source. The parametrisation (\ref{3.1}) is motivated 
by hydrodynamical models with approximately boost-invariant longitudinal 
dynamics. It uses thermodynamic and hydrodynamic parameters and 
appropriate coordinates; for a detailed discussion see, e.g., \cite{H96}. 
In the calculations we use the on-shell approximation $K_0 \approx E_K = 
\sqrt{m^2 + {\bf K}^2}$ as discussed in \cite{CSH95b}.

The velocity field $u(x)$ determines the dynamics of the source at
freeze-out. We parametrize it by \cite{WSH96}
\begin{equation}
 \label{3.3}
   u^{\mu}(x) = \left( \cosh \eta \cosh \eta_t(r), \,
                     \cos \phi \, \sinh \eta_t(r),  \,
                     \sin \phi \, \sinh \eta_t(r),  \,
                     \sinh \eta \cosh \eta_t(r) \right) ,
 \end{equation}
thereby implementing a boost-invariant longitudinal flow profile
$v_L = z/t$, with a linear radial profile of strength $\eta_f$
for the transverse flow rapidity\footnote{The nonrelativistic 
approximation of this transverse profile used in \cite{CSH95b,CL96a,CNH95} 
is advantageous for analytic manipulations but not necessary if the 
correlation radii are evaluated numerically.}:
 \begin{equation}
 \label{3.4}
  \eta_t(r) = \eta_f {r \over R} \, .
 \end{equation}
Parameterizing $K$ in the usual way through rapidity $Y$, transverse 
mass $M_\perp$ and transverse momentum $K_\perp$, the exponent of the 
Boltzmann term reads
  \begin{equation}
    \label{3.5}
    K\cdot u(x) = M_\perp \cosh(\eta - Y) \cosh \eta_t(r) 
                - K_\perp \, \sinh \eta_t(r) \,\cos \phi \,.
  \end{equation}
For $\eta_f = 0$ the emission function depends only
on $M_\perp$ and not on $K_\perp$. This scaling is broken
by the transverse flow.

A special feature of the model suggested in \cite{CL96a} is the 
parametrisation of the temperature profile:
\begin{equation}
\frac 1{T(x)} = \frac 1{T_0} \left ( 1 + a^2\frac {r^2}{2\, R^2} \right )
 \left ( 1 + d^2 \frac { {(\tau - \tau_0)}^2}{2\, \tau_0^2} \right )\, .
\label{3.6}
\end{equation}
It introduces transverse and temporal temperature gradients which are 
scaled by the parameters $a$ and $d$. Such a profile concentrates the 
production of particles with large $M_\perp$ near the symmetry axis and 
close to the average freeze-out time \cite{CL96a,CL96b}. Note that the 
space-time dependence of the temperature does not break the 
$M_\perp$-scaling of the emission function in the absence of transverse 
flow.

\begin{figure}[ht]
\begin{center}
\epsfxsize=12cm
\centerline{\epsfbox{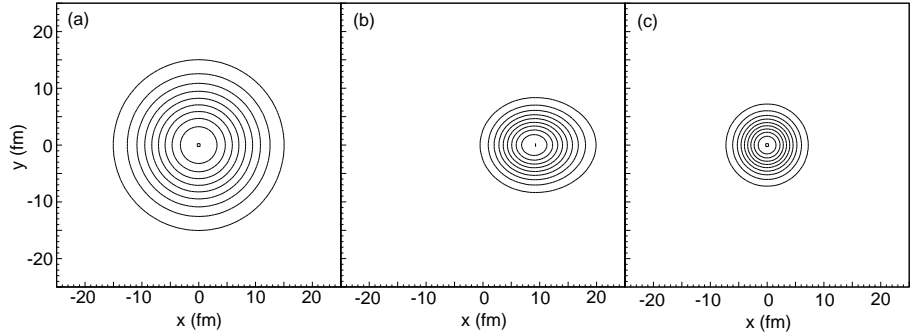}}
\caption{Transverse cuts of the emission function for midrapidity
  pions with transverse momentum $K_\perp=500$ MeV/$c$ (in
  $x$-direction), for the model parameters given in
  Table~\ref{tab1}.
  Left panel (a): neither flow nor temperature gradients;
  middle panel (b): transverse flow only with $\eta_f = 0.7$; right
  panel (c): transverse temperature gradient only with $a=0.8$
}
\label{f1}
\end{center}
\end{figure}

\begin{figure}[ht] 
 \begin{center}
 \epsfxsize=12cm
 \centerline{\epsfbox{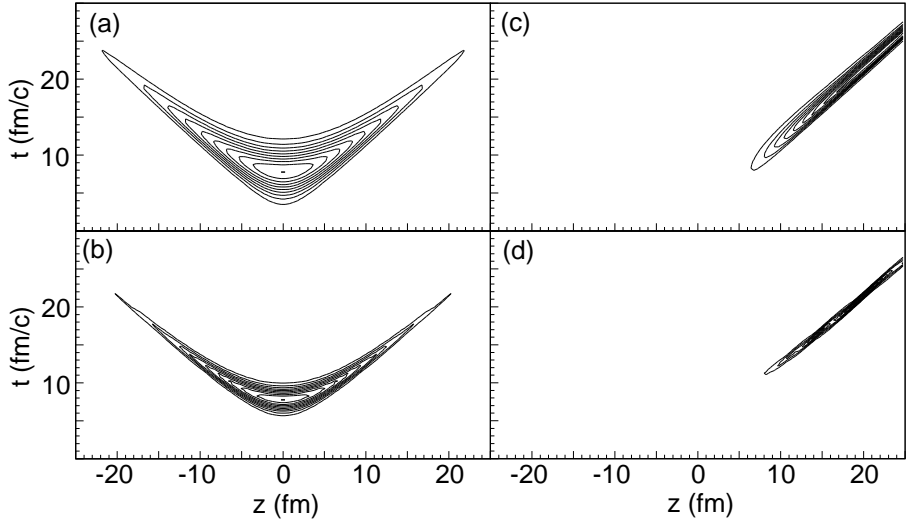}}
 \caption{Contour plots of the $z-t$ profiles of the emission
   function for pions. The profiles are calculated in the CMS. (a,b):
   pair rapidity $Y = 0$, transverse pair momentum $K_\perp = 1$
   MeV/$c$. (c,d): $Y = 2$, $K_\perp =  500$ MeV/$c$. The sensitivity
   of the profiles to $K_\perp$ is weak. Source parameters as given in
   Table~\ref{tab1}, $\eta_f = a = 0$. Upper row (a,c): no temporal
   temperature gradient, $d=0$; lower row (b,d): $d=1.5$}
\label{f2}
\end{center}
\end{figure}

The emission functions (effective sources for pions with given momenta) 
for different model parameters are shown as density contour plots in 
Figs.~\ref{f1} and \ref{f2}. Transverse cuts of the emission function
are displayed in Fig.~\ref{f1}. 
  
Transverse flow is seen to decrease the effective source more in 
the sideward than in the outward direction (see Fig.~\ref{f1}b 
and also Fig.~3 in \cite{WSH96}). Transverse temperature 
gradients, on the other hand, just reduce the homogeneity lengths
without changing the shape of the source (cf. Figs.~\ref{f1}a,c).

A temporal temperature gradient has no impact on the transverse 
source profile; its effect on the longitudinal profile is seen in
Fig.~\ref{f2} (top vs. bottom row) where it decreases the temporal
width of the effective sources without otherwise changing the shape
of the emission function.

Figure~\ref{f2} also shows the different shapes of the effective
source in the center of mass system (CMS) of the fireball for 
midrapidity and forward rapidity pions. An interesting feature
are the emission time distributions: one clearly sees that
the boost-invariant longitudinal expansion and proper-time freeze-out
leads to freeze-out at different times in different places and thus
generates time distributions in a fixed coordinate system which cover 
a much larger region \cite{WSH96,HTWW96a} than the one corresponding 
to the intrinsic eigentime width encoded in the emission function 
(\ref{3.1}). The strongly non-Gaussian shape of the longitudinal
source distribution is known \cite{WSH96} to lead to non-Gaussian
effects on the longitudinal correlator; as shown in \cite{reson1} 
its Gaussian shape is, however, to some extent restored by resonance 
decay contributions (not considered here) which fill in the central 
region above the ``hyperbola'' in the thermal source shown in 
Figs.~\ref{f2}a,b.\footnote{We thank to Urs Wiedemann for making this 
point.}

For our numerical model study we use, if not stated otherwise, 
the source (\ref{3.1}) with the model parameters listed in
Table \ref{tab1}.
 \begin{table}
 \caption{Values of model parameters used in numerical calculations
         }
 \begin{center}
 \begin{tabular}{lc}
 \hline \hline 
  temperature $T_0$ & 100 MeV \\
  average freeze-out proper time $\tau_0$ & 7.8 fm/$c$ \\  
  mean proper emission duration $\Delta \tau$ & 2 fm/$c$ \\
  geometric (Gaussian) transverse radius $R$ & 7 fm \\
  Gaussian width of the space-time rapidity profile $\Delta \eta$ & 1.3 \\
 \hline
  pion mass $m_{\pi^\pm}$ & 139 MeV/$c^2$ \\  
  kaon mass $m_{K^\pm}$   & 493 MeV/$c^2$ \\
 \hline  \hline
 \end{tabular}
 \end{center}
 \label{tab1}
 \end{table}

\section{Influence of temperature and flow gradients on the \\ HBT radii
of a transparent source}
\label{transparent}

In this section we discuss the effects of various types of gradients
in the source on the correlation radii. To be able to recognize their 
specific signals we first investigate the correlation radii in the 
absence of transverse flow and temperature gradients. At the end of 
this section we try with the help of temperature gradients to 
reproduce the $1/\sqrt{M_\perp}$ scaling proposed in \cite{CL96a} 
for $R_s$ and the emission duration. 

For the sake of clarity, let us list here the various reference frames
which will appear in the following discussion. They differ by their 
longitudinal velocities.
 \begin{description}
 \item[CMS] Center of Mass System of the fireball. In this frame $\eta_0 = 0$.
 \item[LCMS] Longitudinally Co-Moving System -- a frame moving longitudinally
        with the particle pair, i.e., $Y = 0$, $\beta_l=0$.
 \item[YK] The Yano-Koonin frame. It moves longitudinally with the 
        Yano-Koonin velocity $v$. The YKP radii measure the homogeneity 
        lengths of the source in this frame. 
 \item[LSPS] Longitudinal Saddle-Point System. This frame moves
        longitudinally with the point where the emission function
        for particles with a given momentum $K$ has its maximum 
        (point of maximal emissivity). 
 \end{description}
It is obvious that each particle pair (via its momentum $K$) defines 
its own LCMS, YK, and LSPS frame. The rapidity of the 
pair in the CMS will be denoted by $Y_{_{\rm CM}} = Y - \eta_0$.

\subsection{No temperature and flow gradients}
\label{nograd}

\begin{figure}[ht]
 \begin{center}
 \epsfysize=10.64cm
 \centerline{\epsfbox{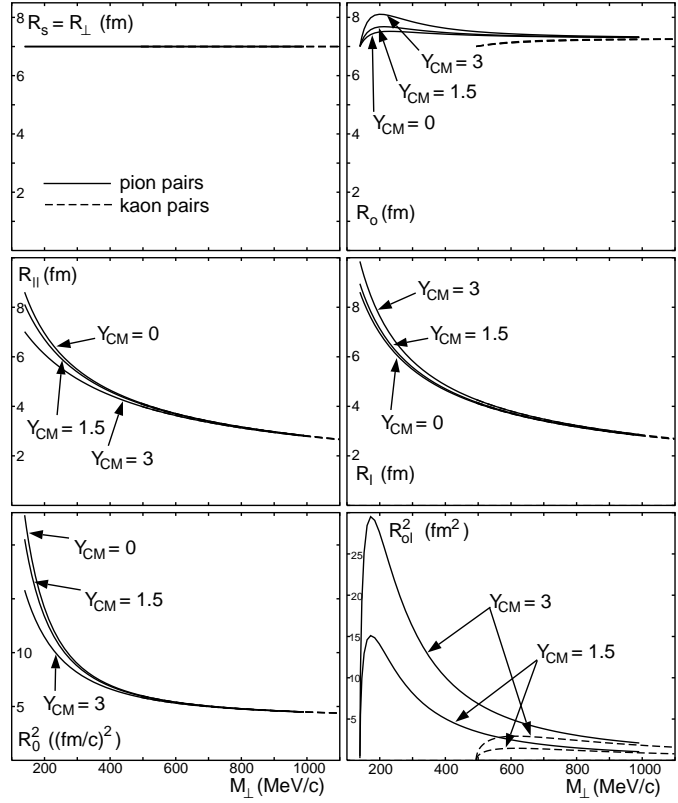}}
 \caption{Correlation radii in the absence of temperature gradients 
   and transverse flow, calculated for a transparent source with  
   model parameters from Table~\ref{tab1}. 
   In the left (right) column the $M_\perp$ dependences of the
   YKP (Cartesian) radii are shown. The space-time variances from
   which they are calculated are evaluated in the LCMS. The HBT radii
   for pions (kaons) are shown as solid (dashed) lines, in three 
   rapidity regions: $Y_{_{\rm CM}} = 0$, $Y_{_{\rm CM}} = 1.5$
   and $Y_{_{\rm CM}} = 3$}
 \label{f3}
 \end{center}
 \end{figure}

In Fig.~\ref{f3}
we show the $M_\perp$ dependence of the correlation radii for both 
the Cartesian and Yano-Koonin-Podgoretski\u\i \ parametrisations,
for a source without transverse flow and temperature gradients.
Non-trivial $M_\perp$ dependencies thus results solely from the
effects of longitudinal expansion or have a kinematic origin.
All HBT radii are calculated from space-time variances of the
emission function evaluated in the LCMS. These curves are shown for
later reference only; for a detailed interpretation of their
features we refer to the existing literature 
\cite{MS88,WSH96,CNH95,HTWW96a,WHTW96,CSH95b}. Here we concentrate on
a few relevant features:

Due to the absence of transverse gradients, $R_s=R_\perp$ is
independent of $K$; it is not affected by longitudinal gradients 
\cite{WHTW96}. The outward Cartesian radius parameter $R_o$ reflects 
the effective lifetime of the source (in the frame in which the
expressions on the r.h.s. are evaluated) via \cite{WSH96,CP91} 
 \begin{equation}
  R^2_{\rm diff} \equiv  R_o^2 - R_s^2 = \beta_\perp^2 \, 
  \left \langle {\tilde t}^2 \right \rangle + 
  2\, \beta_\perp \, \left \langle \tilde x \tilde t \right
  \rangle + \left \langle {\tilde x}^2 - {\tilde y}^2 \right \rangle \,.
 \label{4.1}
 \end{equation} 
Due to the absence of transverse flow the last two terms vanish here.
The effective lifetime in the YK frame (i.e. in the rest frame of the
effective source) is measured by $R_0$ in the YKP parametrisation 
\cite{HTWW96a,WHTW96}. Due to different kinematic factors the
lifetime effects are stronger and more easily visible in $R_0$ than
in $R_o$. Figure~\ref{f2} also confirms the observation \cite{WSH96,HTWW96a}
that the $M_\perp$ dependence of the effective emission duration is 
connected with the $M_\perp$ dependence of the longitudinal region 
of homogeneity which results from longitudinal expansion \cite{MS88}.

The decrease of $R_\parallel$ near the edge of the rapidity
distribution results from an interplay between the Gaussian 
space-time rapidity distribution and the Boltzmann factor in
the emission function. It gives rise to a narrowing rapidity width 
of the effective source with increasing pair rapidity $Y_{_{\rm CM}}$.
At large $M_\perp$, the longitudinal radii $R_l$ and $R_\parallel$
are independent of the pair rapidity $Y_{_{\rm CM}}$, due to the 
longitudinal boost-invariance of the velocity profile in the Boltzmann
factor which dominates the shape of the emission function in the limit
$M_\perp \to \infty$.

It is interesting to compare the longitudinal Cartesian radius 
parameter $R_l$ in the LCMS frame with the longitudinal YKP radius
$R_\parallel$. For small $M_\perp$, $R_l$ is larger than
$R_\parallel$, although not by much, while at large $M_\perp$ the
two parameters agree. This can be understood by recalling the
relation \cite{HTWW96a}
 \begin{equation}
 \label{2.19b}
   R_l^2 = \left(1-\beta_l \right)^2 R_\parallel^2 
         + \gamma^2 \left(\beta_l -v\right)^2 
                    \left(R_0^2 + R_\parallel^2 \right)
 \end{equation}
which, together with two other such relations \cite{HTWW96a}, 
expresses the mathematical equivalence of the Cartesian and YKP
parametrisations. For large $M_\perp$ the source velocity $v$
coincides with the longitudinal velocity $\beta_l$ of the pair
(which is zero in the LCMS). Then the second term in (\ref{2.19b})
vanishes and $R_l^2=R_\parallel^2$ in the LCMS. For smaller values of
$M_\perp$ and $Y_{_{\rm CM}} \ne 0$ the pair and source velocities 
$\beta_l$ and $v$ are slightly different \cite{HTWW96a}; the resulting
positive contribution from the second term in (\ref{2.19b}) renders 
$R_l^2 > R_\parallel^2$ unless $R_0^2$ turns negative (cf.~\cite{th2}). 
This feature was already observed by 
Podgoretski\u\i\ \cite{P83} who introduced the Yano-Koonin frame 
as the frame in which the production process is reflection symmetric
with respect to the longitudinal direction. Within a class of 
non-expanding models he showed that in this frame the longitudinal 
source radius is minimal. On first sight this appears to contradict
the laws of special relativity from which one might expect that
the longitudinal homogeneity length of the source should be {\em largest} 
in the source rest frame and appear Lorentz {\em contracted} in any 
other frame. This argument neglects, however, the fact that different 
points of the homogeneity region generally freeze out at different 
times\footnote{We thank J\'an Pi\v{s}\'ut for a clarifying
discussion on this point.} (see Figs.~\ref{f2}c,d).

The cross term $R_{ol}^2$ shown in the lower right panel of
Fig.~\ref{f3} is required by the invariance of the correlation 
function under longitudinal boosts of the measurement frame 
\cite{CSH95a,CSH95b}. Its value and $M_\perp$ dependence 
depends strongly on the measurement frame; for example, in the CMS
the cross term is negative with smaller absolute value \cite{uliqm96}
although its generic $M_\perp$ dependence remains similar.

Except for $R_o$ and $R_{ol}^2$, all radius parameters shown in
Fig.~\ref{f3} scale with the transverse mass $M_\perp$ and show no
explicit dependence on the particle rest mass. The curves for pions
and kaons thus coincide. This scaling is broken for $R_o$ and
$R_{ol}^2$ by the appearance of explicit factors $\beta_\perp$ in 
Eqs.~(\ref{2.9b}) and (\ref{2.9d}). A quantitative measure for the
strength of the $M_\perp$ dependence can be obtained by fitting
the radius to a power law, $R_i(M_\perp) \propto M_\perp^{-\alpha_i},
\  i=l,\parallel,0$ \cite{AlberPhD,Alber95}. The corresponding
exponents are listed in Table \ref{tab2}.
 \begin{table}
 \caption{The exponents found by a fit of the HBT radii with the
   function $R_i(M_\perp) \propto M_\perp^{-\alpha_i},\  i=l,
   \parallel,0$, for a transparent source without transverse expansion
   and temperature gradients}
 \begin{center}
 \begin{tabular}{||cl|ccc||}
 \hline \hline
 \ & \ & \multicolumn{3}{c||}{$ Y_{_{\rm CM}}$} \\
 \cline{3-5}
 \ & \ & 0 & 1.5 & 3 \\ 
 \hline
 \ & pions & 0.568 & 0.585 & 0.626 \\
 \raisebox{1.5ex}[0ex]{$\alpha_l$} & kaons & 0.546 & 0.553 & 0.573 \\
 \hline 
 \ & pions & 0.568 & 0.542 & 0.478 \\
 \raisebox{1.5ex}[0ex]{$\alpha_\parallel$} & kaons & 0.546 & 0.538 & 0.515 \\
 \hline
 \ & pions & 0.403 & 0.380 & 0.321 \\
 \raisebox{1.5ex}[0ex]{$\alpha_0$} & kaons & 0.202 & 0.197 & 0.185 \\
 \hline \hline 
 \end{tabular} 
 \end{center}
 \label{tab2}
 \end{table}
The longitudinal radius parameters $R_l$ and $R_\parallel$ scale 
approximately with $1/\sqrt{M_\perp}$ as predicted in~\cite{MS88}.
The scaling power is generally closer to -0.5 for kaons than for
pions; this is due to their larger rest mass which reflects in
larger values for $M_\perp$ where the saddle point approximation and 
the assumption of longitudinal boost-invariance in~\cite{MS88} 
become better. For the temporal YKP parameter $R_0$ the
$1/\sqrt{M_\perp}$ scaling law does not hold -- it decreases more slowly,
especially at larger $M_\perp$ and for kaons.       

\subsection{Effects from individual types of gradients}
\label{gradeff}

In this subsection we investigate separately the effects of specific
types of gradients in the emission function on the two-particle
correlations, discussing at the same time the corresponding single
particle spectra. A simultaneous analysis of one- and two-particle
spectra is required for a clear separation of thermal and collective
features of the source and for a complete reconstruction of its
emission function \cite{CL96a,CN96,HTWactual}. 

\begin{figure}[ht]
 \begin{center}
 \epsfysize=6.3cm
 \centerline{\epsfbox{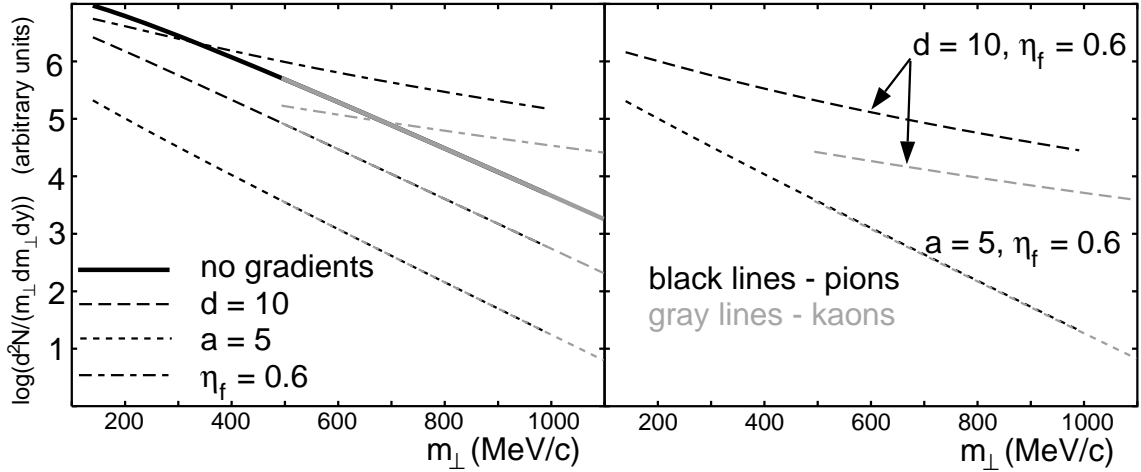}}
 \caption{Transverse mass spectra at midrapidity from a transparent
   source. Black lines correspond to pions, gray lines to kaons. In 
   the left plot the effects from different types of gradients are
   shown individually, in the right plot we study the interplay of
   transverse temperature gradients with transverse flow } 
 \label{f4}
 \end{center}
 \end{figure}

We begin with the discussion of gradient effects on the single
particle spectra. Figure~\ref{f4} 
shows the influence of transverse and temporal temperature gradients 
and of transverse flow on the $m_\perp$-spectra at fixed 
$Y_{_{\rm CM}}=0$. Qualitatively very similar features are seen at  
other rapidity values and in the rapidity integrated spectra.
One sees that even very strong temporal and transverse temperature
gradients do not cause any major effects on the single particle
spectra, compared to the case without any gradients. Their main effect
is a change of the normalisation because the colder regions of the
emission function contribute less; their contribution is also
concentrated at smaller momenta, leading to a somewhat steeper slope
of the transverse mass spectra at low $m_\perp$.

This situation changes dramatically when the source develops
transverse collective flow. In the absence of temperature gradients,
transverse flow simply flattens the $m_\perp$-spectra \cite{SSH93}.
For large $m_\perp \gg m_0$ (i.e. relativistic momenta) this can be
described in terms of an effective blueshifted temperature $T_{\rm eff}
= T \sqrt{{1+\langle v(r)\rangle \over 1 - \langle v(r) \rangle}}$
which is the same for all particle species while at small $m_\perp$
the flattening is even stronger and depends on the particle mass via
the nonrelativistic relation $T_{\rm eff} = T + m \langle v(r)
\rangle^2$ \cite{SSH93}. These features have now been clearly 
observed in the heavy collision systems studied at the Brookhaven AGS
and CERN SPS \cite{qm96}. Transverse flow clearly also breaks the
$m_\perp$ scaling between pions and kaons: the pion and kaon spectra
in the left panel of Fig.~\ref{f4} have different slopes and
normalisations. 

Additional temperature gradients which are imposed on top of the
transverse expansion flow affect the spectra as shown in the right
panel of Fig.~\ref{f4}. A temporal gradient of the source
temperature has no qualitative effect on the shape of the $m_\perp$
spectra and only affects their normalisation. A transverse spatial
gradient of the temperature, however, interferes seriously with the
transverse flow by strongly reducing its effect on the slope of the
$m_\perp$ spectra. For a strong transverse temperature gradient with
$a=5$ as shown in the Figure, a comparison with the left panel shows
that the flow effects become nearly invisible. An analytical
discussion of this behaviour is given in \cite{CL96a} where it is
shown to occur if the transverse homogeneity length generated by
temperature gradient becomes much smaller than the one generated by
the transverse velocity gradient. It follows from this discussion that
it is not possible to generate a strong (additional)
$M_\perp$-dependence of the transverse HBT radius parameter from
transverse temperature gradients (see below)  without at the same time
reducing the flow effects on the single particle $m_\perp$ spectra.

\begin{figure}[ht]
 \begin{center}
 \epsfysize=10.16cm
 \centerline{\epsfbox{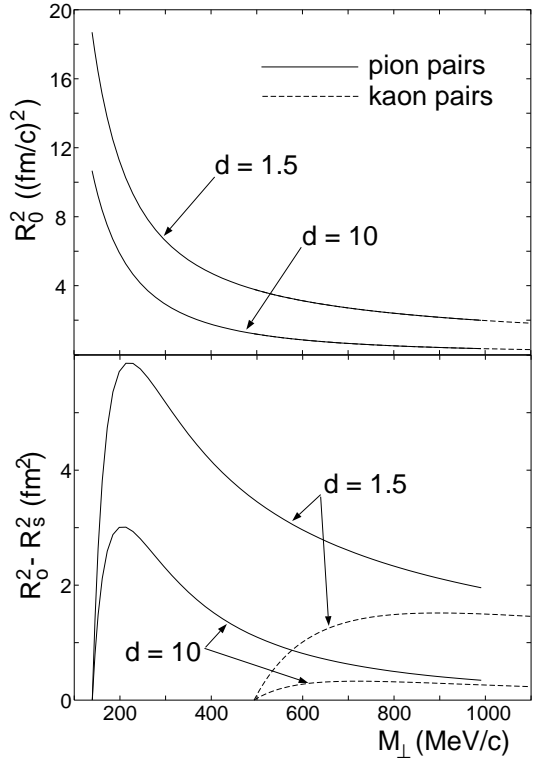}}
 \caption{The effects of temporal temperature gradients on the
   temporal YKP parameter $R_0$ (upper panel) and the difference 
   $R^2_o - R^2_s$ in the Cartesian parametrisation (lower panel).
   Solid (dashed) lines are for pions (kaons). The calculations are
   done for midrapidity pairs, $Y_{_{\rm CM}} = 0$, and for
   $a = \eta_f = 0$}
 \label{f5} 
 \end{center}
 \end{figure}

We now proceed to a discussion of the HBT radius parameters. The
effect of temporal temperature gradients is studied in Fig.~\ref{f5}.  
As expected, they affect mostly those HBT parameters which are
sensitive to the effective lifetime of the source, namely the temporal
YKP parameter $R_0$ and the Cartesian difference $R_o^2 - R_s^2$. 
These are reduced by increasing $d$. The origin of this effect can be 
seen in Fig.~\ref{f2} which shows that larger values of $d$
concentrate the high temperature region (with large emissivity) to a
narrower region in the $z$-$t$ plane. The narrowing occurs
predominantly in the temporal direction, but affects slightly also the
longitudinal homogeneity length. Going from central rapidity to pairs
at forward rapidities, all curves change in a similar way as shown 
in Fig.~\ref{f3}.

\begin{figure}[ht]
 \begin{center}
 \epsfysize=6.6cm
 \centerline{\epsfbox{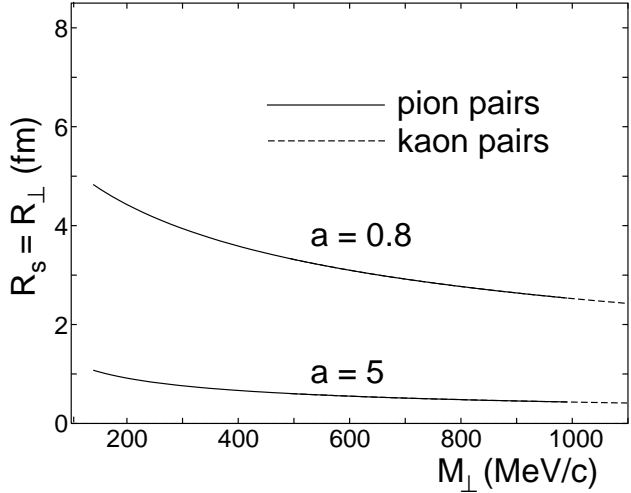}}
 \caption{The effect of a transverse temperature gradient: The
   transverse radius parameter $R_s=R_\perp$ as a function of the
   transverse mass of the pair, for two different values of $a$. The
   coincidence of the solid (dashed) lines for pions (kaons) reflects
   the $M_\perp$-scaling of the transverse radius in this case. $d =
   \eta_f = 0$,  $Y_{_{\rm CM}} = 0$} 
 \label{f6}
 \end{center}
 \end{figure}

Transverse temperature gradients leave their traces only in the
transverse homogeneity length measured by $R_s = R_\perp$. In
Fig.~\ref{f6} 
this is shown for pions and kaons at mid-rapidity. (Note that in our model 
$R_s$ is rapidity independent.)

The most important feature of both types of temperature gradients is that 
they {\em do not break the $M_\perp$ scaling of the YKP correlation radii} 
found in previous subsection. Note also that the temperature gradients
have no qualitative impact on the YK rapidity $Y_{_{\rm YK}}$
associated with the YK velocity $v$.

\begin{figure}[ht]
 \begin{center}
 \epsfxsize=9.45cm
 \epsfysize=10.64cm
 \centerline{\epsfbox{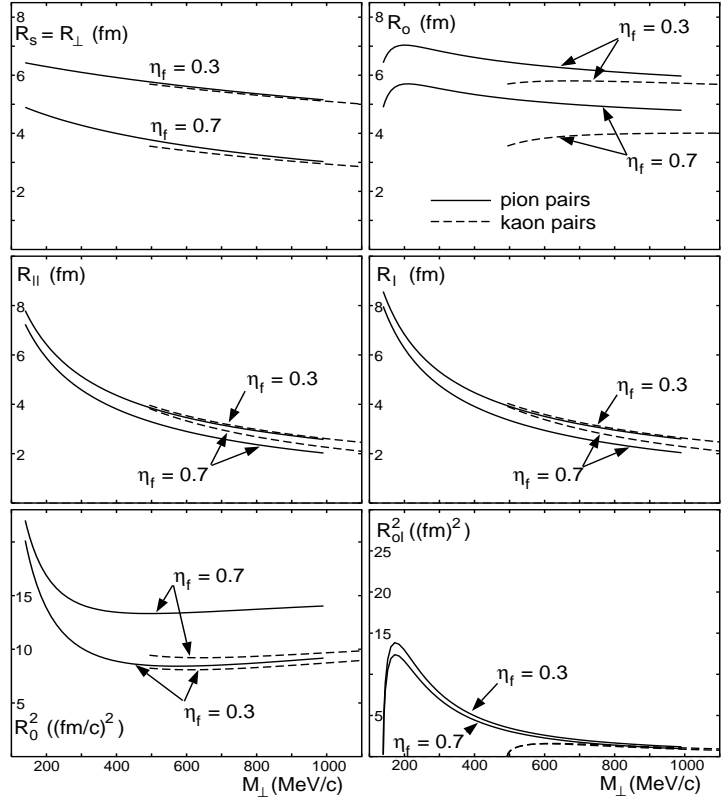}}
 \caption{The correlation radii for a transparent source of constant
   temperature featuring transverse collective flow. The rapidity of
   the pairs was chosen as $Y_{_{\rm CM}} = 1.5$. Left column: YKP
   radius parameters. Right column: Cartesian radius parameters.
   Solid (dashed) lines correspond to pions (kaons)}
 \label{f7}
 \end{center}
 \end{figure}

Many of these features change in the presence of transverse flow. This
is shown in Fig.~\ref{f7} 
for pairs with rapidity $Y_{_{\rm CM}}=1.5$ in the CMS. All radius
parameters have been calculated from space-time variances evaluated in
the LCMS. The choice of a non-zero CMS pair rapidity ensures a
non-vanishing cross term $R_{ol}$. The generic rapidity dependence of
the radii follows the behaviour discussed in Sec.~\ref{nograd}. 

Since transverse flow introduces velocity gradients in the transverse 
direction, $R_s ( = R_\perp)$ and $R_o$ decrease with increasing
$\eta_f$. Less obvious is, however, the rather strong effect of
transverse flow on the ``temporal'' YKP parameter $R_0^2$. For higher
transverse masses the latter even begins to increase with
$M_\perp$. This does not, however, reflect the behaviour of the
emission duration which continues to decrease for particles with
higher $M_\perp$. The increase of $R_0^2$ is rather caused by
correction terms expressing the difference between $R_0^2$ and the
lifetime \cite{HTWW96a,WHTW96}: 
 \begin{equation}
  R_0^2 - \langle {\tilde t}^2 \rangle  = 
  \frac {2}{\beta_\perp}  \langle \tilde x \tilde t
   \rangle + \frac 1{\beta_\perp^2}  \langle
  {\tilde x}^2 - {\tilde y}^2  \rangle \, .
 \label{4.2}
 \end{equation}
Especially the second term on the r.h.s. grows appreciably with
increasing $\eta_f$, and even more so for pions than for kaons. This
was studied in detail in~\cite{WHTW96} (cf. Fig.~2 of that paper)
where a more detailed explanation of the curves for $R_0^2$ shown here
can be found. 

The most important feature of Fig.~\ref{f7} is the {\em breaking of the
  $M_\perp$ scaling of the YKP radii by the transverse flow}.

\subsection{Interplay of all gradients}
\label{interplay}

In~\cite{CL96a} it was claimed that an interplay of the various
types of gradients discussed above can lead to a common 
$1/\sqrt{M_\perp}$ scaling law for all three Cartesian HBT radii,
$R_l$, $R_s$, and $R_o$, as well as for the effective lifetime of the
source. The condition for such a common scaling is that the
geometric extension is large compared to the ``thermal length scales''
generated by temperature and flow gradients \cite{CL96a}. The claim 
of~\cite{CL96a} was based on approximate analytical expressions for
these radius parameters which were obtained by evaluating the
corresponding integrals over the emission function within an improved
saddle-point approximation scheme. In this subsection we present a
numerical test of this claim. To this end we select a set of
parameters ($a = 0.6$, $d = 10$, $\eta_f = 0.2$) for which at
mid-rapidity the conditions for such a scaling, as given in~\cite{CL96a}, 
are satisfied as much as possible, without leaving
the phenomenologically realistic range. It should be mentioned,
that our model differs slightly from that of~\cite{CL96a}. The differences
are explicitly treated in Appendix~\ref{tam_model} where they are  
shown to be small, and our findings are not affected by them.

\begin{figure}[ht]
 \begin{center}
 \epsfysize=10.75cm
 \centerline{\epsfbox{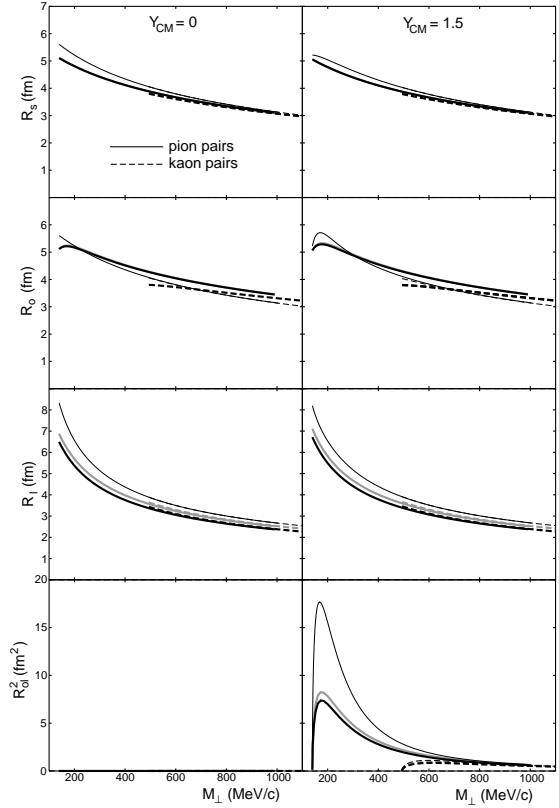}}
 \caption{Correlation radii in the Cartesian parametrisation for a
   transparent source with parameters $a = 0.6$, $d = 10$, $\eta_f =
   0.2$. The thick black lines represent the exact
   result from a numerical evaluation of the space-time variances, 
   the thick grey lines the result of numerical evaluation with the
   model of~\cite{CL96a} with corresponding parameters 
   (see Appendix~\ref{tam_model}), the
   thin lines use the approximate analytical formulae given in
   \protect\cite{CL96a}. Solid (dashed) lines are for pions (kaons).
   In cases of $R_s$ and $R_o$ thick black and grey lines overlap}
 \label{f8}
 \end{center}
 \end{figure}

The numerically evaluated correlation radii for this parameter set 
are plotted with thick black lines in Fig.~\ref{f8}.
The analytical approximation according to~\cite{CL96a} is shown
by the thin lines for comparison. With thick grey lines the 
numerically computed results  from original model of~\cite{CL96a} 
introduced in Appendix~\ref{tam_model} are plotted.
We agree with the conclusions drawn
from a similar comparison presented in~\cite{CLL96} that for
kaons the approximate analytical formulae agree with the numerical
curves within 15\%. For pions with transverse momenta below 600 MeV
the discrepancies are larger because one of the conditions of validity
of the analytical formulae (sufficiently large $M_\perp$) is violated.
However, the analytical approximation does not give at all the (slight)
breaking of the $M_\perp$ scaling of $R_l$ by the transverse flow, and
at mid-rapidity it misses the initial rise of $R_o$ at small
$M_\perp$. This last point in particular is a serious problem if one
wishes to extract an estimate of the effective lifetime according to 
Eq.~(\ref{4.1}). Similar (dis-)agreement at a comparative level is
seen for forward rapidity pairs at $Y_{_{\rm CM}}=1.5$, with the
exception of the cross term $R_{ol}$ which is strongly overestimated
by the approximation. The analytical approximation of~\cite{CL96a} 
is restricted by the condition that the flow
rapidity of the point of maximum emissivity in the LCMS is
small. Since the latter coincides with good approximation with the YK
rapidity \cite{HTWW96a} and, in our model, the YK rapidity is found to
converge to $Y$ in the limit $M_\perp \to \infty$
\cite{HTWW96a,WHTW96}, the point of maximum emissivity moves to zero
LCMS rapidity in that limit. This explains why for large $M_\perp$ the
analytical and numerical results agree.

On a superficial level the analytical approximations of~\cite{CL96a} 
thus don't seem to be doing too badly.
We want to check, however, whether the results also confirm 
the suggested  $1/\sqrt{M_\perp}$-scaling. In~\cite{WHTW96} 
it was shown that in general the strength of the
$M_\perp$-dependence of $R_s$ and $R_l$ (resp. $R_\perp$ and
$R_\parallel$) is correlated with the strength of collective
flow in the transverse resp. longitudinal directions. One way of
quantifying the strength of the $M_\perp$-dependence of the
correlation radii is to fit them to a power law $R_i \propto
M_\perp^{-\alpha_i} \ (i = s,  l,  \parallel , 0)$ and to study
the values of the exponents $\alpha_i$. According to the approximate
formulae of~\cite{CL96a}, at mid-rapidity all these powers should
be equal to 0.5. The $\alpha_i$-values obtained from the numerically
evaluated correlation radii are listed in Table~\ref{tab3}.
 \begin{table}
 \caption{The exponents $\alpha_i$ found in a fit of the numerically
   computed correlation radii to a power law $R_i \propto M_\perp^{-
     \alpha_i}$, for a source with the parameters given in Fig.~\ref{f8}}
\begin{center}
\begin{tabular}{||cl|ccc||}
\hline \hline
\ & \ & \multicolumn{3}{c||}{$ Y_{_{\rm CM}}$} \\
\cline{3-5}
\ & \ & 0 & 1.5 & 3 \\ 
\hline 
\  & pions &  0.257 & 0.253 & 0.240 \\
\raisebox{1.5ex}[0ex]{$\alpha_s$} & kaons & 0.306 & 0.304 & 0.297 \\
\hline
\ & pions & 0.512 & 0.527 & 0.565 \\
\raisebox{1.5ex}[0ex]{$\alpha_l$} & kaons & 0.519 & 0.524 & 0.538 \\
\hline
\ & pions & 0.512 & 0.496 & 0.453 \\
\raisebox{1.5ex}[0ex]{$\alpha_\parallel$} & kaons & 0.519 & 0.512 & 0.494 \\
\hline
\ & pions & 0.373 & 0.360 & 0.324 \\
\raisebox{1.5ex}[0ex]{$\alpha_0$} & kaons & 0.225 & 0.223 & 0.262 \\
\hline \hline
\end{tabular} 
\end{center}
 \label{tab3}
 \end{table}
One sees that only the longitudinal radius parameters (both in YKP and
Cartesian parametrisations) follow an approximate $1/\sqrt{M_\perp}$
scaling law, similar to previous studies without temperature
gradients \cite{WSH96,WHTW96}. It reflects the boost-invariance of the
longitudinal expansion velocity profile \cite{MS88}. The transverse
and temporal radius parameters $R_s=R_\perp$ and $R_0$ scale much more
weakly with $M_\perp$. For
mid-rapidity kaons $R_0$ provides a good estimate for the emission
duration because for $\eta_f=0.2$ the correction terms on the
r.h.s. of (\ref{4.2}) are small; thus we find that also the 
effective lifetime of the source does not scale with
$1/\sqrt{M_\perp}$. Thus, while the conditions for the saddle point
approximation in~\cite{CL96a} are satisfied with sufficient
accuracy by our (semi-realistic) parameter set, the stronger
conditions required for the common $1/\sqrt{M_\perp}$-scaling are not.


\section{Conclusions}
\label{conclusions}

Let us shortly summarize the most important results:

Transverse temperature gradients and transverse flow have similar 
effects on the transverse HBT radii $R_s=R_\perp$, but not on the 
single particle spectra. Both lead to a decrease of $R_s$ with 
increasing $M_\perp$, but while transverse flow flattens the single 
particle spectra, a transverse temperature gradient only reduces 
their normalisation. A weak $M_\perp$-dependence of $R_s$ due to a 
moderate transverse flow can be {\em increased} by adding a transverse 
temperature gradient, but only at the expense of simultaneously {\em 
reducing} the flow effects on the single particle spectra. This is 
important for the interpretation of experimental data.

For transparent sources, transverse temperature gradients preserve the 
$M_\perp$-scaling of the YKP radius parameters while transverse flow 
breaks it. This breaking of $M_\perp$-scaling is weak, however, the 
most sensitive parameter being $R_0^2$.  

Temporal temperature gradients affect mostly the temporal HBT radii, 
i.e. the difference $R_o^2-R_s^2$ in the Cartesian parametrisation and 
$R_0^2$ in the YKP parametrisation.

We could not confirm the existence of a common $1/\sqrt{M_\perp}$
scaling law for all three Cartesian HBT radius parameters and the 
effective source lifetime in the case of the ``thermal lengths'' 
being smaller than  the geometric lengths.
For {\em realistic} model parameters, the $M_\perp$-dependence for $R_s$ and 
$R_0$ is always much weaker than for the longitudinal radius parameters 
$R_l$ resp. $R_\parallel$.

\vspace{4ex}

{\bf Acknowledgements:} 
We thank to Tamas Cs\"org\H o for very fruitful discussions 
which helped us to formulate the model study and comparison
with analytical formulae more accurately. We are also indebted 
to Harry Appelsh\"auser,  J\'an Pi\v{s}\'ut, Claus Slotta, 
Axel Vischer and Urs Wiedemann for valuable and helpful comments.  
We acknowledge financial support by DAAD, DFG, BMBF, and GSI.

\appendix
\section{The model of Cs\"org\H o and L\"orstad}
\label{tam_model}

In this Appendix we point out the differences between our model 
and that of~\cite{CL96a} and study their impact on our findings 
from Sec.~\ref{interplay}. 

The model of \cite{CL96a} is given by the emission function
\begin{eqnarray}
	S(x,K)\, d^4x & = & \frac{g}{(2\pi)^3} M_\perp \cosh(\eta-Y)\,
	\exp\left[ - \frac{K\cdot u(x)}{T(x)} + \frac{\mu(x)}{T(x)}\right]
	\nonumber \\ & & \times \;
	\frac{\tau_0^\prime \, d\tau}{\sqrt{2\pi\, {\Delta\tau^\prime}^2}}
	\exp\left[- 
	\frac{(\tau-\tau_0^\prime)^2}{2\, {\Delta\tau^\prime}^2}\right]
	\, d\eta\, r\, d\phi\, dr \, .
\label{b1}
\end{eqnarray}
Since in the used coordinates the measure reads 
$d^4x = d\tau\, \tau \, d\eta\, r \, d\phi\, dr$,
it is seen that here another geometric eigentime distribution 
is used, namely
\begin{equation}
	H(\tau) = \frac{\tau_0^\prime}{\tau}\, 
 	\frac1{\sqrt{2\pi\, {\Delta\tau^\prime}^2}}
	\exp\left[- 
	\frac{(\tau-\tau_0^\prime)^2}{2\, {\Delta\tau^\prime}^2}\right] \, .
\label{b2}
\end{equation}
This differs from our distribution by the pre-factor 
${\tau_0^\prime}/{\tau}$. For the comparison with our model, 
the values of ${\tau_0^\prime}$ and 
$\Delta \tau^\prime$ are obtained via the saddle point approximation
to our Gaussian eigentime distribution multiplied by $\tau$ from the 
Jacobian. This gives
\begin{eqnarray}
	\tau_0^\prime & = & \frac{\tau_0 + \sqrt{\tau_0^2 + 4\Delta\tau^2}}2
	\, ,\label{b3} \\
	\Delta\tau^\prime & = & \frac{\tau_0^\prime\, \Delta\tau}{
	\sqrt{\Delta\tau^2 + {\tau_0^\prime}^2}} \, .
	\label{b4}
\end{eqnarray}
The remaining geometry of the model (\ref{b1}) is encoded in 
chemical potential
\begin{equation}
\label{b5}
	\frac{\mu(x)}{T(x)} = \frac{\mu_0}{T_0} - \frac{r^2}{2\, R^2}
	- \frac{(\eta-\eta_0)^2}{2\, \Delta \eta^2} \, ,
\end{equation}
so this is the same as in  our model~(\ref{3.1}). There is a slight difference 
in the formulation of temperature gradients:
\begin{equation}
\label{b6}
	\frac1{T(x)} = \frac1{T_0} \left( 1 + {a^\prime}^2
	\frac{r^2}{2\, {\tau_0^\prime}^2} \right)\left( 1 +
	{d^\prime}^2 \frac{(\tau - \tau_0^\prime)^2}{2\, {\tau_0^\prime}^2}
	\right)\, ,
\end{equation}
where the transverse gradient scales with ${a^\prime}/{\tau_0^\prime}$
(and not with $a/R$). Furthermore, the transverse flow rapidity 
is given as
\begin{equation}
\label{b7}
	\eta_t = \eta_f^\prime \frac{r}{\tau_0^\prime}\, ,
\end{equation}
instead of $\eta_f \, r/R$. This requires appropriate re-scaling of 
$a^\prime$ and $\eta_f^\prime$. Also, $d^\prime$ has to be re-scaled
relatively to $d$ by the factor $\tau_0^\prime/\tau_0$. Note finally
that in the numerical calculation we have used the expansion four-velocity
field as given by the relativistic formula~(\ref{3.3}) with the transverse
rapidity profile (\ref{b7}), while in \cite{CL96a} the non-relativistic
approximation to the transverse expansion is used in the formulation of 
the model.

The results are plotted with grey lines in Fig.~\ref{f8}. It is clearly
seen that the difference between the mentioned two models is small. 

For the computation of the curves using the analytical formulae 
found in \cite{CL96a}
(thin lines in Fig.~\ref{f8})
we used the re-scaled model parameters as explained above.

To be in the region where the $1/\sqrt{M_\perp}$-scaling should take place
we have chosen: $a = 0.6$, $d = 10$, $\eta_f = 0.2$. 
Note that with the appropriately re-scaled parameter
values (see above) we satisfy the
conditions given in~\cite{CLL96} which need to be satisfied for 
a 15-20\% agreement between the numerical calculation and the
analytical approximations of the model (\ref{b1}).

To be sure that the slope parameters found in Table~\ref{tab3} are
not an artefact of the difference between our model and that of \cite{CL96a}
we did the same fit with the numerically calculated radii resulting
from~(\ref{b1}) and even with the analytically computed curves. Results of 
that fit are listed in Table~\ref{tab-tamas}. 
 \begin{table}
 \caption{The exponents $\alpha_i$ found in a fit of the numerically
   computed correlation radii to a power law $R_i \propto M_\perp^{-
     \alpha_i}$, for a source from \protect\ref{b1} 
     and with the parameters correspondingly re-scaled to approximate 
     those used in  Fig.~\ref{f8}. In brackets are the results of the 
     same fit done with curves computed using the analytical
     approximation from~\cite{CL96a} }
\begin{center}
\begin{tabular}{||cl|lll||}
\hline \hline
\ & \ & \multicolumn{3}{c||}{$ Y_{_{\rm CM}}$} \\
\cline{3-5}
\ & \ & 0 & 1.5 & 3 \\ 
\hline 
\  & pions &  0.257 (0.301) & 0.253 (0.274) & 0.240 \\
\raisebox{1.5ex}[0ex]{$\alpha_s$} & kaons & 0.306 (0.370) & 0.304 (0.364)
      & 0.297 \\
\hline
\ & pions & 0.511 (0.569) & 0.526 (0.562) & 0.564 \\
\raisebox{1.5ex}[0ex]{$\alpha_l$} & kaons & 0.518 (0.531) & 0.523 (0.535)
     & 0.538 \\
\hline
\ & pions & 0.511 & 0.495 & 0.452 \\
\raisebox{1.5ex}[0ex]{$\alpha_\parallel$} & kaons & 0.518 & 0.512 & 0.493 \\
\hline
\ & pions & 0.395 & 0.381 & 0.344 \\
\raisebox{1.5ex}[0ex]{$\alpha_0$} & kaons & 0.245 & 0.242 & 0.279 \\
\hline \hline
\end{tabular} 
\end{center}
 \label{tab-tamas}
 \end{table}
The differences of  the
exponents listed in Tables~\ref{tab3} and \ref{tab-tamas} 
are in case of $\alpha_s$, $\alpha_l$ and $\alpha_\parallel$ within 1\%, 
$\alpha_0$'s differ up to 10\%. 
The qualitative discussion 
concerning the exponents from Sec.~\ref{interplay} 
remains valid for the set given in Table~\ref{tab-tamas}. The same is true for the analytically determined
radii. 

This again supports the conclusion, that even if good agreement between
numerical and analytical results is achieved, for a realistic set of 
parameters the conditions for common scaling of all three Cartesian 
radii with $M_\perp^{-1/2}$ are not fulfilled.

\begin{flushleft}

\end{flushleft}

\end{document}